\documentclass[11pt]{article} 
 
\usepackage{amssymb,cite} 
\usepackage{graphicx} 
\usepackage{latexsym} 

\usepackage{epsfig,amsfonts} 

\usepackage{bm}
\usepackage[ansinew]{inputenc}
\usepackage{rotating}
\usepackage{euscript}
 
\def\thefootnote{\fnsymbol{footnote}}

\newcommand{\pathC}{\mathcal C}

\newcommand{\fd}[1]{\left[\mathcal{D}#1\right]}
 
\newcommand{\eq}{\begin{equation}} 
\newcommand{\en}{\end{equation}} 
\newcommand{\be}{\begin{equation}} 
\newcommand{\ee}{\end{equation}} 
\newcommand{\eqa}{\begin{eqnarray}} 
\newcommand{\ena}{\end{eqnarray}} 
\newcommand{\ba}{\begin{eqnarray}} 
\newcommand{\ea}{\end{eqnarray}}

\newcommand{\bra}{\left \langle}
\newcommand{\ket}{\right \rangle}

\newcommand{\ZZ}{\hbox{{\rm Z{\hbox to 3pt{\hss\rm Z}}}}} 
\def\de{\partial}

\newcommand{\EQ}{\begin{equation}} 
\newcommand{\EN}{\end{equation}} 
\newcommand{\bea}{\begin{eqnarray}} 
\newcommand{\eea}{\end{eqnarray}}

\newcommand{\myfrac}[2]{\frac{\quad\displaystyle #1\quad}{\quad\displaystyle #2\quad}}

\begin{document} 
\begin{titlepage} 
\vskip0.5cm 
\begin{flushright} 
\end{flushright} 
\vskip0.5cm 
\begin{center} 
{\Large\bf  Flux tube delocalization at the deconfinement point.} 
\end{center} 
\vskip1.3cm 
 
\centerline{M. Caselle}    

 \vskip0.4cm    
 \centerline{\sl  Dipartimento di Fisica    
 Teorica dell'Universit\`a di Torino}
 \centerline{\sl  and} 
 \centerline{\sl INFN, Sezione di Torino,}    
 \centerline{\sl via P.Giuria 1, I-10125 Torino, Italy}    
 \centerline{\sl    
e--mail: \hskip 1cm caselle@to.infn.it}    
 \vskip1cm    
    
\begin{abstract}    
We study the behaviour of the flux tube thickness in the vicinity of the deconfinement transition. We show, using effective string methods,
that in this regime the square width increases linearly and not logarithmically with the interquark distance. 
The amplitude of this linear growth is an increasing function of the temperature and diverges  
as the deconfinement transition is approached from below. These predictions are
in good agreement with a set of simulations performed
in the 3d gauge Ising model. 

\end{abstract}    
\end{titlepage}    
 
\setcounter{footnote}{0} 
\def\thefootnote{\arabic{footnote}} 
\section{Introduction} 
\label{introsect}

One of the most intriguing features of the confining regime of Lattice Gauge Theories (LGTs) 
is the logarithmic increase of the square width $w^2(R)$ of the
 flux tube as a function of the interquark distance $R$~\cite{lmw80}. 
This effect, which is commonly referred to as the  "delocalization" of the flux 
tube is
together with the well known L\"uscher term one of the most important and stringent
 predictions of the effective string approach to LGTs.

This effect was discussed for the first time many years ago by 
 L\"uscher, M\"unster and Weisz in~\cite{lmw80} but it required several years of efforts before it could be observed 
 in lattice simulations. The first numerical results were obtained in abelian models ~\cite{cgmv95, Zach:1997yz, Koma:2003gi, Giudice:2006hw}
 and only recently they were extended also to non abelian LGTs~\cite{Gliozzi:2010zv,Bakry:2010zt} (see also ~\cite{Bali:1994de} for some early attempt). 
 In all these tests a good agreement with the theoretical predictions was found.  
 
 A question which naturally arises in this context is which is the fate of the flux tube (and in particular the behaviour of its width) 
 as the deconfinement transition is approached from below. It is important to stress that delocalization and deconfinement are two deeply different conditions of the flux
 tube. The deconfinement transition is characterized by the vanishing of the string tension $\sigma(T)$. In pure gauge theories it is associated to the breaking of the
 center symmetry of the gauge group and its natural order parameter is the Polyakov loop. At the deconfinement point the flux tube vanishes. 
 The delocalization of the flux tube instead coincides with the roughening transition. In all the
 physically interesting models this transition occurs for value of the gauge coupling $\beta$ well below the fixed point where the continuum limit can be taken. Thus in
 the continuum limit (for any temperature below the deconfinement transition) the theory is always in the rough phase and thus the flux tube is always delocalized. 
 Delocalization is a typical quantum
 effect. It is a consequence of the Mermin-Wagner theorem which imposes the restoring (in the continuum limit) 
 of the translational symmetry for the fluctuations of the flux tube in the transverse directions. Intutively it amounts to say that we cannot fix deterministically 
 the trajectory of the flux tube but may only describe it as a probabilty distribution.
 It is important to stress that, even if
 delocalized, the flux tube fully keeps its confining function. The quantum fluctuations which drive the delocalization also influence the confining potential
 (as the presence of L\"uscher term indicates) but do not destroy it. 
  
 While the behaviour of the string tension $\sigma(T)$ as the deconfinement temperature $T_c$ is approached from below is rather well understood much less is 
 known on the behaviour of the flux tube thickness in this regime.
 This is a rather important issue from a physical point of view 
 since the interplay between delocalization and deconfinement could strongly influence the transition from hadrons to free quarks as $T_c$ is approached.  
 
 This problem can be addressed 
 within the framework of the effective string approach by performing a modular 
 transformation of the low temperature result. This was done in~\cite{Allais:2008bk} in the case of the free bosonic approximation (i.e. the first order in the perturbative
 expansion of the Nambu-Goto effective string).  
 In this free bosonic limit one can show  that when periodic boundary conditions are imposed in the time direction (i.e. when finite temperature regularization 
 is imposed) the large $R$ behaviour of the square width changes
 completely and becomes linear instead of logarithmic. This behaviour holds in principle for any temperature $T$, but as $T$ decreases it requires larger and larger values
 of $R$ to be observed. Similarly it is possible to show that for any fixed value of $R$  the square width smoothly converges toward the expected 
 logarithmic behaviour as $T$ decreases. The threshold between the two
 behaviours is  $R\sim 1/T$. The coefficient of the linear dependence at high $T$ is itself temperature dependent. It 
 increases linearly with the temperature and remains finite as the deconfinement transition is approached.
  
 In~\cite{Allais:2008bk} these predictions were compared with a set of high precision simulations in the 3d gauge Ising model, with results which turned out to be only in 
 partial agreement with the effective string picture. For all the temperatures studied in~\cite{Allais:2008bk} $w^2(R)$ was indeed, 
 (with very good confidence levels) a linearly increasing function of $R$. 
 However the coefficient of this linear behaviour was in general larger
 than that predicted by the effective string (except for the smallest temperature values) 
 and, what is more important, it seemed to diverge as the deconfinement point was approached.
 
 It is clear that in order to understand this discrepancy one should go beyond the first order approximation 
 in the effective string calculation. As a matter of fact in this finite temperature description, the perturbative expansion in powers of $1/\sigma_0 L^2$,
 is actually an expansion in $T^2/T_c^2$ (see below for further details on this point) 
 thus it is somehow obvious that higher orders terms become important if one wants to study the behaviour in the vicinity of the
 deconfinement point. What is less obvious is how relevant are for the flux tube width these higher order contributions.
 Indeed it seems that the particular problem that we are addressing here, the high temperature behaviour
 of the flux tube thickness, is one of the best numerical laboratories to test higher order contributions in the effective string description. 
 In fact  the
 comparison with the 3d Ising model~\cite{Allais:2008bk} shows that these higher order corrections may become very large (much larger than the first order term) 
 in the vicinity of the deconfinement transition.
 
 Higher order effective string corrections have attracted much interest in these last years both as a tool to better fit the numerical results and as a way to 
 improve our understanding of the interplay between the effective string approach and standard string theory. 
 However no result had been obtained until very recently on higher order corrections to the flux tube width. 
 Most of the papers concentrated instead on higher order
 corrections to the interquark potential. Several important results were obtained in this context. It was shown that not only the leading (L\"uscher) term
 but also the first subleading (quartic) correction is universal~\cite{lw04,Drummond:2004yp,HariDass:2006sd}, thus remarkably
 enhancing the predicting power of the effective model. In a recent paper universality has been extended also to the subsequent term (sixth order term 
 in the potential)~\cite{Aharony:2009gg}
 at least for the (2+1) dimensional case and for the torus and cylinder topologies. Finally, in the case of the Nambu-Goto string, the all order expansion was obtained 
 in~\cite{lw04,bc05}. 
  
 The situation changed recently thanks to~\cite{Gliozzi:2010zv} where 
 an explicit expression for the next to leading (quartic) correction to the flux tube width was obtained in
  the case of the cylinder topology. This is exactly the topology which is needed to describe the behaviour of the flux tube at finite temperature.

  The aim of the present paper is to use the result of ~\cite{Gliozzi:2010zv} to address the discrepancy between theoretical predictions and numerical results 
   described above. As we shall see, keeping into account this correction the right $T$ dependence of the flux tube thickness is recovered.
   This, together with a parallel result  obtained using a dimensionally reduced effective model for the high temperature behaviour of lattice gauge theories,
   will allow us to guess further terms in the expansion and to propose the following conjecture for 
   the behaviour for the large $R$ limit of the flux tube square width up to the deconfinement transition:
\eq 
\sigma(T)w^2(R)\, = \,  \frac{1}{4} \, {R}{T} 
\label{sy3bis}
\en 
where $\sigma(T)$ denotes the temperature dependent string tension.

If, in addition, one also assumes a Nambu-Goto effective action for the flux tube (assumption which we know to be wrong, but in several cases gives a very  good
approximation, at least for the first few orders in $T$) then the $T$ dependence of  $\sigma(T)$
can be predicted exactly and turns out to be  
\eq 
  \sigma(T)=\sigma_0  
  \left(1-\frac{T^2}{T_c^2} 
  \right)^{1/2}. 
\label{sigmat}
  \en 
Thus allowing a complete description of the $T$ dependence of the flux tube thickness in the limit of large interquark separation. 
      
\section{Effective string prediction for the flux tube thickness at finite temperature} 
 
\subsection{Definition of the flux tube thickness} 

In a finite temperature setting the lattice operator which is used to evaluate the flux through a plaquette $p$ of the lattice is:
\eq 
\bra\phi(p;P,P')\ket=\frac{\bra P P'^\dagger~U_p\ket}{\bra PP'^\dagger \ket}-\bra U_p\ket 
\label{flux2} 
\en 
where $P$, $P'$ are two Polyakov loops separated by $R$ lattice spacings and $U_p$ is 
the operator associated with the plaquette $p$.
Different possible orientations of the
plaquette $p$ measure different components of the flux. In the following we shall neglect this dependence which
plays no role in our analysis. The only information that we need is the position of the plaquette.
Let us define 
\[
 \bra\phi(p;P,P')\ket=\bra\phi(\vec h;R,L)\ket
\]
where $\vec h$ denotes the displacement of $p$ from the $P$ $P'$ plane. 
In each transverse direction, the flux density 
shows a gaussian like shape (see for instance Fig. 2 of ~\cite{cgmv95}). 
The width of this gaussian $w$ is the quantity which is usually denoted as ``flux tube thickness'':

\eq
w^2(R,L)=\frac{\sum_{\vec h} \vec h^2 \bra\phi(\vec h;R,L)\ket}{\sum_{\vec h} \bra\phi(\vec h;R,L)\ket}
\label{w1}
\en

This quantity depends on the number of transverse dimensions and on the bare gauge coupling $\beta$ (or, equivalently, on the lattice spacing $a$). 
Once $\beta$ is fixed the only remaining dependences are 
on the interquark distance $R$ and  on the lattice size in the compactified timelike 
direction $L$, i.e. on the inverse temperature of the model. 
By tuning $L$ we can thus study the flux tube thickness near the deconfinement transition.

\subsection{Effective string prediction: general setting}
In the effective string  framework the square of flux tube width is given by:
\eq
  w^2(x;R,L)=\myfrac
  {\int_{\pathC}\fd{h}h^i(t,x)h^i(t,x)e^{-S[h]}}
  {\int_{\pathC}\fd{h}e^{-S[h]}}
\label{w2}
\en
where the sum is intended over all the surfaces bordered by the two Polyakov loops 
and $S[h]$ is the effective action. As mentioned above, up to the second order in which we are interested here, this action is universal 
and coincides with the expansion to the fourth order in $h$ of the Nambu-Goto action. In the following we shall concentrate in the (2+1) dimensional case (i.e. we shall
deal with only one transverse direction) and eliminate the index of the field $h$. In this case the
Nambu Goto action has a particularly simple form
\eq
S[h]=\sigma_0\int_{-L/2}^{L/2}d\tau\int_{-R/2}^{R/2} d\varsigma\sqrt{1+(\de_\tau h)^2+(\de_\varsigma h)^2} \;\;.
\label{squareroot}
\en
The correlator eq.(\ref{w2}) is singular and must be regularized. The standard choice is the point splitting regularization.
It is easy to see looking at eq.(\ref{squareroot}) that, in the large $R/L$ limit in which we are interested 
there is a natural expansion parameter 
in the evaluation of eq.(\ref{w2}) which is $1/(\sigma_0L^2)$.
The leading order in this expansion is simply given by the free bosonic approximation to the Nambu-Goto action while the next to leading correction
will include the fourth order terms of the action. Let us address these two contributions in more detail. 

\subsection{Effective string prediction: leading order}
The leading order effective string prediction for square of the flux tube thickness $w^2_{lo}$ in the finite temperature (i.e. cylinder) 
geometry was evaluated for the first time (in a rather implicit form) in~\cite{cgmv95}. It was then reobtained in a slightly more explicit form in 
\cite{Allais:2008bk} using the method of images. We shall use as our starting point this last expression:
 \eq
\begin{array}{c}
\displaystyle
\sigma_0 w^2_{lo}=-\frac{1}{2\pi} \log\frac{ \pi |\epsilon|}{2 R} + 
\frac{1}{2\pi} \log\left|\theta_2(0)/\theta_1'(0) \right|
\end{array}
\label{ris1}
\en
where $\sigma_0$ denotes the zero temperature string tension and
$L$ is the length of the cylinder in the compactified direction (i.e. the inverse temperature)
and the square width is evaluated at the midpoint between the two Polyakov loops.

It is easy to see using standard relations among $\theta$ and $\eta$ functions that this expression 
coincides with the one reported in~\cite{Gliozzi:2010zv}.

In the large $R$ and high temperature limit, i.e. for $R>>L$ this expression becomes:

\eq
  \sigma_0 w^2_{lo}=\frac{1}{2\pi} \log\frac{L}{L_c} +
  \frac{R}{4L}-\frac{1}{\pi}e^{-2\pi\frac{R}{L}}+\cdots
\label{risfinale}
\en 

which shows, as anticipated, that the square width should increase linearly with $R$ with a 
coefficient $\frac{1}{4\sigma_0 L}$ which increases linearly with the temperature and stays finite at the deconfinement point. 
In fact the zero temperature string
tension $\sigma_0$ (which can be extracted from the Wilson loop or from Polyakov loop correlators at very low temperature)
does not depend on the finite temperature $T$ but only on the coupling constant $\beta$.

\subsection{Effective string prediction: next to leading correction}

The next to leading correction was recently evaluated in~\cite{Gliozzi:2010zv}.We report here for completeness the result: 

\begin{eqnarray}
\label{widthcorr}
&&w^2_{nlo}=\left( 1 +\frac{4\pi f(\tau)}{\sigma_0 r^2}\right) w^2_{lo}
(r/2)-\frac{f(\tau)+g(\tau)}{\sigma_0^2r^2}, \nonumber \\
&&f(\tau)=\frac{E_2(\tau)-4E_2(2\tau)}{48}, \nonumber \\
&&g(\tau)=i \pi \tau \left(\frac{E_2(\tau)}{12}- \frac {qd}{dq} \right) 
( f(\tau)+\frac{E_2(\tau)}{16} )+\frac{E_2(\tau)}{96}, 
\end{eqnarray}
where $E_2(\tau)$ is the first Eisenstein function: 

$$E_2(\tau)=1-24\sum_{n=1}^\infty\frac{n\,q^n}{1-q^n}.$$

$\tau\equiv \frac{i L}{2 R}$ denotes the modulus of the cylinder and
$$q=\mbox{e}^{2 \pi i \tau}=\mbox{e}^{ -\pi L/R}$$

The dominant term in  the  $R>>L$ limit (i.e. $\tau \to 0$) turns out to be again linear in $R$:

\eq
   w^2_{nlo}\sim
  \frac{\pi R}{24\sigma_0^2 L^3}+\cdots
\label{risnlo}
\en 

This result is rather non trivial. Looking at the small $\tau$ expansion of the Eisenstein function 
one would in principle expect a term proportional to $R^2$ which however disappears in eq.(\ref{widthcorr}) as a
consequence of a set of non trivial cancellations among Eisenstein functions. This is an important consistency check in view of the fact that the simulations reported 
in~\cite{Allais:2008bk} found no evidence of terms proportional to $R^2$ and found instead a remarkably precise linear increase of $w^2$ with $R$.

Combining together the leading and subleading corrections we end up with the following expression:

\eq
   w^2=  \frac{R}{4\sigma_0 L}\left(1+
  \frac{\pi}{6\sigma_0 L^2}\right)+\cdots
\label{rislast}
\en 

which, assuming the Nambu-Goto value for the critical temperature $T_c^2=3 \sigma_0/\pi$
(see below for a derivation of this result), can be rewritten as an expansion in $T/T_c$

\eq
   w^2=  R\, \frac{3}{4\pi}\frac{T}{T_c^2}\left(1+
  \frac{T^2}{2T_c^2}\right)+\cdots
\label{rislast2}
\en

\begin{figure}[htpb]
  \centering
  \includegraphics[width=6 in]{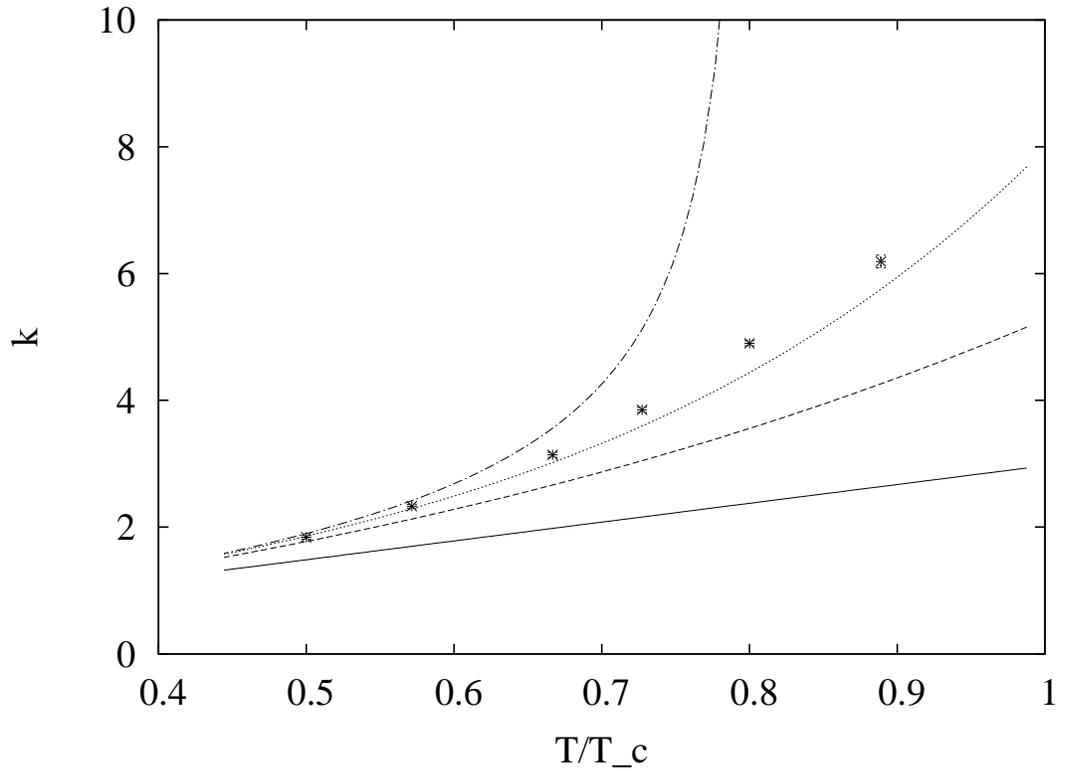}
  \caption{\small \textit{Plot of $k(L)$ as a function of $\frac{T}{T_c}=\frac{8}{L}$. The continuous line is the prediction for 
  $k(L)$ according to eq.(\ref{risfinale}). The dashed line according to eq.(\ref{rislast}). The dashed dotted line to eq.(\ref{sy3}) and
  the dotted line to (\ref{sy4}). In all the four cases the curves were obtained only using the terms explcitly written in the corresponding equations
  and neglecting higher order corrections (denoted with the dots in the various equations).
   The points are the results of the simulations in the 3d gauge Ising model
  }}
  \label{fig1}
\end{figure}

We compare in tab.\ref{tab1} and fig.\ref{fig1} (dashed line) 
this result with the values of the numerical simulations reported in~\cite{Allais:2008bk}. We see a clear improvement with respect to the
leading contribution alone (continuous line), which however is not
enough to fill the gap with the numerical data.
At the same time it is easy to see that, as anticipated, for values of $L$ in the vicinity of the deconfinement point the 
subleading term becomes of the same order of the leading one. This shows that in this limit the expansion in powers of $1/\sigma_0 L^2$  converges too slowly and that
it would be very important to have some educated guess on how to resum the expansion. We shall show in the next section that this guess can be obtained with a
completely different approach, i.e. looking at the effective theory obtained integrating out the spacelike degrees of freedom following the well known 
Svetitsky-Yaffe proposal~\cite{sy82}.

\subsection{Dimensional reduction and the Svetitsky Yaffe approach.} 
In the vicinity of the deconfinement transition the physics of a (d+1) LGT can be described using an effective model in which the spacelike links
are integrated out and the only remaining degrees of freedom are the Polyakov loops. This essentially amounts to integrate out all the degrees of freedom
of the original LGT up to the scale $L=1/T$ and in fact  
the theory obtained in this way effectively behaves as a model in one dimension less than the original LGT. It is easy to see that this procedure gives as effective
description a
$d$ dimensional spin model (the Polyakov loops playing the role of spins) with a global symmetry represented by the center $C$ of the gauge group of the original LGT. 
This approach was pioneered by Svetitsky and Yaffe~\cite{sy82} who were able to predict in this way the critical behaviour of the LGT at the deconfinemenet point under the
assumption that both the deconfinement transition of the original LGT and the magnetization transition of the effective spin model are continuous. However its utility 
goes beyond the prediction of the critical indices. In the last few years it has been shown that this effective description can be used 
to predict the behaviour of various quantities in the vicinity of the deconfinement transition as far as the scales involved in these observables (for
instance the interquark distance $R$) are larger than $L=1/T$ which sets the scale of the effective theory (for a review of most 
recent results see for instance~\cite{Caselle:2009yv}). 
This is exactly the $R>>L$ limit in which we are interested in
this paper.

The simplest examples of this effective mapping are 
the (2+1) $SU(2)$ LGT  and the (2+1) Ising gauge model which have the same center $Z_2$ and are thus both mapped into the 2d spin Ising model.
In this case, since both LGTs have a
second order deconfinement phase transition we can also predict, following~\cite{sy82}, that these transitions are in the same universality class of the
2d Ising model. 

Let us address in more detail these examples. It is easy to see that the confining phase of the LGT is mapped 
into the high temperature phase of the spin model and that,
as we mentioned before, the Polyakov loops of the LGT are mapped into the spins of the 2d Ising  
model. This correspondence should be understood in the renormalization group sense
i.e.  the Polyakov loop operator is actually mapped into a linear combination of all
the (C-odd) operators of the spin model. In the 2d Ising case this means that we have a combination involving
the whole conformal family of the spin operator. For $R$ large enough this combination is
dominated by the relevant operators  which in the Ising case is only the
spin operator. In a similar way it is possible to show that the plaquette operator is mapped into the most general 
combination of the energy and the identity conformal families. This allows us to construct the analogous of the operator which measures the flux tube thickness
which turns out to be a suitable combination of three point correlators of the spin and energy operators (see~\cite{Caselle:2006wr} for a detailed
discussion of this mapping). In the particular case of the 2d Ising model these correlators can be evaluated exactly leading to the following  expression for
the "flux"  distribution~\cite{Caselle:2006wr}
\eq 
P(R,y) \ = \frac{ 2\pi  R}{4y^2+R^2} \, \frac{e^{-m \sqrt{4y^2+R^2}}}{K_0(mR)}.  
\label{nongaussian} 
\en 
 where $y$ denotes the transverse direction, $K_0$ is the modified Bessel function of order 0,  $m$ is the mass of the 2d Ising model
and a large $mR$ limit is assumed.

From this flux distribution it is easy to extract the square of the flux tube width as the ratio

\eq 
w^2(R) \ = \frac{\int_{-\infty}^{\infty} dy \, y^2 \, P(R,y)}{\int_{-\infty}^{\infty} dy \, P(R,y)}  
\label{sy1}
\en 
which, setting $x=2y/R$ amounts to evaluate
\eq 
w^2(R) \ = \frac{R^2}{4} \, \frac{\int_{-\infty}^{\infty} dx \, \frac{x^2}{1+x^2} e^{-2mr \sqrt{1+x^2}}}{\int_{-\infty}^{\infty} dx \, \frac{e^{-2mr \sqrt{1+x^2}}}{1+x^2} }  
\en 
These integrals can be evaluated asymptotically in the large $mR$ limit (see~\cite{Caselle:2006wr}\footnote{Notice, to avoid confusion, 
that in \cite{Caselle:2006wr} we used the variable $r\equiv R/2$ and that we evaluated the unnormalized square width, i.e. only the numerator of eq.(\ref{sy1})}) 
leading to the following result:
\eq 
w^2(R) \simeq \  \frac{1}{4} \, \frac{R}{m} + \dots.
\label{sy2}
\en 
where the dots stay for terms constant or proportional to negative powers of $R$.

The last step in order to compare this result with eq.(\ref{rislast}) is to give a meaning to the Ising mass $m$ in terms of LGT quantities.

This can be easily accomplished if we recall that the mass is the inverse of the correlation length of the model and can be obtained from the large $R$ limit of the
2 point function of the model
as follows:
\eq 
\lim_{R\to\infty}  \left\langle \sigma(0,0) \sigma(0,R)\right\rangle 
  \sim  
  K_{0}({m} R). 
\label{rtoinfty1} 
\en 

According to the mapping discussed above this correlator is the 2d limit of the expectation value of two Polyakov loops at distance $R$.
In order to be consistent with the assumptions we made to evaluate the flux tube width in the previous section,  
we must evaluate this correlator in the framework of the 
 Nambu-Goto effective action. This result was obtained a few years ago by L\"uscher and Weisz~\cite{lw04} using a duality transformation 
and then derived in the covariant formalism in~\cite{bc05}. In $d=2+1$ dimensions one finds a tower of $K_0$ Bessel functions: 
\eq 
  \left\langle P(0,0) P(0,R)\right\rangle 
  =\sum_{n=0}^{\infty}c_n 
  K_{0}({E}_nR). 
\en 
(see eq. (3.2) of~\cite{lw04}). 
 
where $c_n$ are constants (whose exact expression is irrelevant for our analysis, the only important point is that they do not contain any $R$ dependence in (2+1)
dimensions, contrary to the higher dimensional cases) and $E_n$ are the closed string energy levels: 
  
 \eq 
  {E}_n=\sigma_0 L
  \left\{1+{8\pi\over\sigma_0 L^2}\left[-\frac{1}{24}\left(d-2\right)+n\right] 
  \right\}^{1/2}. 
  \en 
  (see eq. (C5) of~\cite{lw04}).

It is easy to see that in the large $R$ limit only the lowest state $(n=0)$ survives and, in full agreement with our dimensional reduction mapping,
 we end up with a single 
$K_0$ function: 
\eq 
\lim_{R\to\infty}  \left\langle P(0,0) P(0,R)\right\rangle 
  \sim  
  K_{0}({E}_0R). 
\label{rtoinfty} 
\en 
 
comparing with eq.(\ref{rtoinfty1}) we immediately recognize $m=E_0$ i.e.
\eq 
  m=\sigma_0 L 
  \left(1-{\frac{\pi}{3\sigma_0 L^2}} 
  \right)^{1/2}. 
\label{sigmal}
  \en 
This result can be rewritten as
\eq 
  m=\sigma(T) L 
\en 
where $\sigma(T)$ is given by eq.(\ref{sigmat}) which wew report here:
\eq 
  \sigma(T)=\sigma_0  
  \left(1-\frac{T^2}{T_c^2} 
  \right)^{1/2}. 
\label{sigmatsec}
  \en 
where the critical temperature $T_c^2=3 \sigma_0/\pi$ is the Nambu-Goto prediction for the deconfinement temperature and is given by the solution of the equation
$\sigma(T)=0$.
Eq.(\ref{sigmatsec}) is the prediction of the Nambu-Goto effective action for the
temperature dependence of the string tension
and was discussed several years ago by Olesen in
\cite{Olesen:1985ej}.

Plugging this result in eq.(\ref{sy2}) we find for the coefficient of the linear term:
\eq 
w^2(R) = \  \frac{1}{4} \, \frac{R}{\sigma(T) L} +\dots = \  \frac{1}{4} \, \frac{R}{\sigma_0 L
  \left(1-{\frac{\pi}{3\sigma_0 L^2}} 
  \right)^{1/2}} + \dots.
\label{sy3}
\en 

which may be expanded in powers of $1/\sigma_0 L^2$ as follows

\eq 
w^2(R) =   \frac{1}{4} \, \frac{R}{\sigma_0 L}
  \left(1+\frac{\pi}{6\sigma_0 L^2} +\frac{\pi^2}{24\sigma_0^2 L^4}
  +\cdots\right)
\label{sy4}
\en 
or equivalently
\eq
   w^2=  R\, \frac{3}{4\pi}\frac{T}{T_c^2}\left(1+
  \frac{T^2}{2T_c^2}+ \frac{3 T^4}{8 T_c^4}\right)+\cdots
\label{sy4bis}
\en

The first two orders exactly coincide with those of eq.(\ref{rislast}). This 
represents a remarkable consistency check of both  the Svetitsky-Yaffe approach and the second order perturbative calculation of \cite{Gliozzi:2010zv} 
and leads to conjecture that eq.(\ref{sy3}) could be the resummation to all orders in $1/(\sigma_0 L^2)$ of the linear term of the square width
in the framework of the Nambu-Goto effective action.

It is important to stress, as a final comment on this result, that even if it was obtained assuming  
for $\sigma(T)$ the functional form predicted by the Nambu-Goto action: eq.(\ref{sigmat}) its validity goes well beyond 
this particular model.
In fact the calculations which lead to  eq.(\ref{rislast})
 and to the first two terms of eq.(\ref{sy4bis}) 
only involve contributions up to the next to leading order in the effective string perturbative expansion which were proved to be universal 
in~\cite{lw04,Drummond:2004yp,HariDass:2006sd}.

\begin{table}[htpb]
\centering
\begin{tabular}{|c|c|c|c|c|c|}
  \hline
  $L$ & $k(L)$  & eq.(\ref{risfinale}) & eq.(\ref{rislast}) & eq.(\ref{sy3})  & eq.(\ref{sy4})    \\
  \hline\hline
  9   & 6.19(10) & 2.639 & 4.260  & NAN  & 5.753     \\
  10  & 4.90(4)  & 2.375 & 3.557  &33.307 & 4.438      \\
  11  & 3.85(4)  & 2.159 & 3.047  & 5.121 & 3.594   \\
  12  & 3.14(4)  & 1.979 & 2.663  & 3.560 & 3.017  \\
  14  & 2.33(3)  & 1.697 & 2.127   &2.418  & 2.291 \\
  16  & 1.84(3)  & 1.484 & 1.773   &1.899 & 1.857 \\
  \hline
\end{tabular}
\caption{\small \textit{Values of the coefficient k(L) of eq.(\ref{defk}) 
for various values of $L$. In the second column we list the results of the simulations.
In the following columns we 
report the prediction for $k(L)$ assuuming eq.(\ref{risfinale})(\ref{rislast})
(\ref{sy3})(\ref{sy4}) and neglecting higher order corrections denoted by the dots in the equations.}}
\label{tab1}
\end{table}

\subsection{Comparison with the Ising model.}

In fig.\ref{fig1} and tab.\ref{tab1} we compare eq.(\ref{sy3}) 
with the results of a set of simulations performed in the 3d gauge Ising model and reported 
in~\cite{Allais:2008bk}. In these simulations the coupling of the Ising model was fixed at $\beta=0.75180$ for which the 
deconfinement transition is known to occur  at exactly $8$ lattice spacings~\cite{ch96}. Moreover 
for this value of $\beta$ the zero temperature string tension is known
with very high  precision to be $\sigma_0=0.0105255(11)$~\cite{Caselle:2007yc} thus excluding a possible source of systematic errors. 
The model was  studied for the values of $L$ reported in
the first column of tab.\ref{tab1} which correspond to temperatures ranging from $T=T_c/2$ to $T=8T_c/9$. For each temperature 
 $w^2$ was then evaluated for several values of $R$ and  fitted with
 \eq
 w^2(R,L)=k(L) R + c(L) 
\label{defk}
 \en
For all temperatures the reduced $\chi^2$ of the fits was very good. The values of $k(L)$ obtained in this way are reported in  
the second column of tab.\ref{tab1}.
Looking at fig.\ref{fig1} and tab.\ref{tab1} we see that the for all the temperatures eq.(\ref{sy3}) predicts  values of $k(L)$
larger than those found in the simulations and diverges for $T\sim 4T_c/5$. This is not surprising. Indeed it is by now well known that the
Nambu-Goto effective string is not a good description of the 3d gauge Ising model. In fact it predicts a deconfinement 
temperature which
is  $\sim 0.8$ of the observed one (in agreement with what we find here) and predicts too large corrections in the interquark potential.
This disagreement was discussed in the past years in a set of high precision numerical tests both in the high $T$~\cite{Caselle:2002ah} (but still in the confining
phase) and 
in the low $T$~\cite{Caselle:2004jq} regimes
of the interquark potential 
and more recently looking at the interface free energy of the 3d Ising spin model~\cite{Caselle:2007yc}. This last observable 
is related by duality to the interquark potential and represents an important cross test because it involves different (i.e. toroidal)
boundary conditions. 
In all these tests the interquark potential was rather well described by the truncation at the second order of the
Nambu-Goto expansion in powers of $1/\sigma_0 L^2$ and definitly incompatible with any further higher order contribution.
Remarkably enough we see the same pattern also in this case. We report in fig.1 (the dotted line) and tab.1 (last column) the values of k(L) obtained
truncating eq.(\ref{sy3}) at the second order (i.e. eq.(\ref{sy4})). As shown in fig.1 the measured values lie slightly above this curve. 
This is indeed a puzzling result in view of the recent claim \cite{Aharony:2009gg} on the universality of the sixth order term in the 
effective string expansion for the interquark potential.
This sixth order term seems instead to be very different from the Nambu-Goto one (and compatible this zero) in the case of the 3d Ising model. 
We have for the moment no good answer to this puzzle.

\section {Conclusions}
Let us summarize our main results and add a few concluding comments:
\begin{description}
\item{1)}
As anticipated in the introduction our main result is that approaching the deconfinement transition has a twofold effect on the square width of the flux tube. 
First, the logarithmic dependence on the interquark distance is changed into a linear one. 
Second, (see eq.(\ref{rislast2})) the amplitude of this linear growth increases more than linearly with the temperature as the transition point is
approached.  

\item{2)}
We conjecture that the behaviour described by eq.(\ref{rislast2}) is only the first term of an expansion in powers of $T^2/T_c^2$
 and that the square width $w^2$ should be
proportional to the inverse of the finite temperature string tension (see eq.(\ref{sy3}))
If we isolate in eq.(\ref{sy3})  the term linear in $R$ we find an universal relation which should hold for any LGT in the limit of large interquark distances
\eq 
\sigma(T)w^2(R)\, = \,  \frac{1}{4} \, {R}{T} 
\nonumber
\en 
Moreover, if we assume that the effective action is of the Nambu-Goto type, then we may also predict the explicit expression for $\sigma(T)$ which is
given by  eq.(\ref{sigmat}). 

\item{3)} However it is well know that for many models 
(and in particular for the 3d gauge Ising model)   the Nambu-Goto picture is not correct and that eq.(\ref{sigmat})
is a rather poor description of the actual finite temperature behaviour 
of the string tension $\sigma(T)$. In the particular case of the 3d gauge Ising model we find a good agreement between predictions and numerical data 
if we truncate at the
second order the perturbative expansion in powers of $1/\sigma_0L^2$ in agreement with what already found when looking at the interquark potential.

\item{4)} On the contrary in the case of (2+1) $SU(N)$ gauge models (and in particular for $N=3$) 
the Nambu-Goto string gives a much better description of the finite temperature behaviour of the theory (see \cite{Liddle:2008kk}) and accordingly we
expect that eq.(\ref{sy3}) should well describe the finite temperature behaviour of $w^2$. It would be very interesting to extend to these cases the present analysis, using
for instance the methods of ~\cite{Gliozzi:2010zv} or of~\cite{Bakry:2010zt} and to test on a quantitative basis our prediction eq.(\ref{sy3bis}).

\item{5)} For LGTs with a first order deconfinement transition (as it happens for instance for $N\geq 4$ in $d=2+1$~\cite{Liddle:2008kk} and for $SU(3)$ in $d=3+1$), 
there is no reason to expect a divergent behaviour of the coefficient $k(L)$ . It would be very
interesting to see if eq.(\ref{sy3bis}) still holds in these cases.

\end{description}

\vskip1.0cm {\bf Acknowledgements.}
 I would like to warmly thank  F. Gliozzi, P.Grinza and M.Pepe 
 for useful suggestions and discussions.

\end{document}